\documentclass[%
 reprint,
 amsmath,amssymb,
 aps,
]{revtex4-2}

\usepackage{graphicx}
\usepackage{dcolumn}
\usepackage{bm}

\usepackage{amsmath, amsthm, amssymb, amsfonts,mathrsfs}

\usepackage[linktocpage=true]{hyperref}
\hypersetup{
	colorlinks,
	citecolor=blue,
	filecolor=black,
	linkcolor=blue,
	urlcolor=blue
}
\usepackage{url}
  
\usepackage{physics}

\newcommand{\w}{\omega}
\newcommand{\lz}{\lambda_0}
\newcommand{\uz}{\mu_0}

\newcommand{\ddp}{\delta-\delta'}

\begin{document}

\preprint{APS/123-QED}


\title[Interference phenomena in the asymmetric dynamical Casimir effect for a single \texorpdfstring{$\ddp$}{TEXT} mirror]{Interference phenomena in the asymmetric dynamical Casimir effect\\ for a single \texorpdfstring{$\ddp$}{TEXT} mirror}

\author{Matthew~J.~Gorban}
\email{matthew\_gorban1@baylor.edu}%
\author{William~D.~Julius}
\email{william\_julius1@baylor.edu}%
\author{Ramesh~Radhakrishnan}
\email{ramesh\_radhakrishna1@baylor.edu}%
\author{Gerald~B.~Cleaver}
\email{gerald\_cleaver@baylor.edu}%
\affiliation{%
 Department of Physics, Baylor University,  Waco, TX 76798, USA 
}%

\date{\today}

\begin{abstract}
The interaction between the quantum vacuum and time-dependent boundaries can produce particles via the dynamical Casimir effect. It is known that, for asymmetric Casimir systems, there is an imbalance in the particle production on either side of the boundary. Here, we consider a real massless scalar field in 1+1 dimensions interacting with a moving $\ddp$ mirror with time-dependent properties. The spectral distribution and particle creation rate are computed, which now include an additional interference term that can affect different parts of the spectrum in a constructive or destructive manner. The asymmetry of the system is investigated by analyzing the difference in particle spectra produced on the two sides of the mirror. Additionally, we also explore enhancement of the spectrum and its asymmetry within the context of a stationary $\ddp$ mirror subject to multiple fluctuation sources.
\end{abstract}

\maketitle


\section{\label{sec:intro}Introduction}

A quantized field subjected to time-dependent boundary conditions will interact (exchange energy) with the quantum vacuum to produce real particles in a phenomenon called the dynamical Casimir effect (DCE). Originally introduced by Moore \cite{moore1970quantum}, and expanded upon by the works of DeWitt \cite{dewitt1975quantum}, Fulling and Davies \citep{fulling1976radiation,davies1977radiation}, and Candelas and
Deutsch \cite{candelas1977vacuum}, there is now an abundance of literature on the DCE. See \cite{dodonov2009dynamical,dodonov2010current,dodonov2020fifty} 
for several detailed reviews of this topic. 

As shown by Moore in his pioneering work \cite{moore1970quantum}, there are practical limitations for DCE experiments, as it is difficult to overcome the physical limitations required to mechanically oscillate materials at frequencies on the order of GHz, as is required for measurable particle production \cite{silva2011simple,dodonov2020fifty,nation2012colloquium,wilson2011observation}. While there have been several clever experimental proposals of mechanically induced DCE \cite{kim2006detectability,brownell2008modelling,motazedifard2018controllable,sanz2018electro,qin2019emission,butera2019mechanical}, there are still many challenges to overcome \cite{dodonov2020fifty}. This issue has led to the proposal of alternative methods for observing particle production due to the DCE. Inspired by some of the early work done by Yablonovitch \cite{yablonovitch1989accelerating}, a number of different proposals have been introduced which show that a mirror with time-varying material properties can give rise to the DCE. In particular, the varying material properties will introduce time-dependent boundary conditions in a similar manner to physically oscillating the mirror \cite{schutzhold1998quantum,dodonov1993quantum,braggio2005novel,yablonovitch1989accelerating,kim2006detectability,de2007quantum,gunter2009sub,johansson2009dynamical,wilson2010photon,dezael2010analogue,nation2012colloquium}. Experimental evidence supports the real production of particles from time-varying materials \cite{wilson2011observation,lahteenmaki2013dynamical,schneider2020observation,vezzoli2019optical}. Most notably, the first experimental DCE detection modulated the inductance of a superconducting quantum interference device (SQUID) to alter the electrical length of a superconducting circuit  \cite{wilson2011observation}.

One of the more astounding consequences of quantum vacuum interactions occurs when objects with asymmetric boundary conditions are subjected to time-dependent fluctuations \cite{gorban2023asymmetric,silva2020motion,silva2016dynamical,donaire2016net,donaire2013casimir,donaire2015transfer,feigel2004quantum,croze2013does,croze2012alternative,birkeland2007feigel}. Within the context of the DCE, it is possible to construct a mirror whose surface properties on either side of the mirror are different. This leads to an asymmetric production of particles on either side of the mirror in what is known as the asymmetric dynamical Casimir effect (ADCE) \cite{gorban2023asymmetric}. One consequence of this asymmetric production of particles is that a previously stationary mirror will begin to move due to the unbalanced radiation pressures between the two sides \cite{silva2020motion}.

To model such an asymmetric system, a $\ddp$ potential \cite{barton1995quantum1,nicolaevici2001quantum,nicolaevici2009semitransparency,dalvit2000decoherence,munoz2015delta,braga2016casimir,silva2016dynamical,silva2020motion,rego2022dynamical} (here $\delta$ is the Dirac delta) is used to simulate a partially transparent mirror interacting with the (1+1)-dimensional spacetime, or (1+1)D, quantum vacuum. Thus, the ADCE model becomes (hereafter $c=\hbar=1$)
\begin{equation}\label{lagrangian}
    \mathscr{L}=\frac{1}{2}\big[(\partial_t\phi)^2-(\partial_x\phi)^2\big]-\qty[\mu \delta(x)+\lambda \delta^\prime(x)]\phi^2(t,x),
\end{equation}
where $\mu$ is related to the plasma frequency and $\lambda$ is a dimensionless factor that controls the degree of asymmetry in the system. The inclusion of the $\delta'$ term is what gives rise to the asymmetric boundary interaction, and so when $\lambda=0$ the asymmetry vanishes and the system reduces to a $\delta$ mirror \cite{barton1995quantum1,barton1995quantum2}. This asymmetry manifests in the reflection and transmission coefficients that determine the scattering interactions between the mirror and the vacuum. This asymmetry means that the reflection coefficients on each side of the mirror will not be equivalent. 

The $\ddp$ mirror has been investigated using the standard DCE generation techniques, with both time-varying materials properties \cite{silva2020motion} and fluctuations in mirror position \cite{silva2016dynamical}. Here, we investigate the ADCE for a single moving $\ddp$ mirror that possesses both sources for particle creation. More precisely, we will examine the interaction between the (1+1)D real massless scalar field and the time-dependent $\ddp$ mirror at the instantaneous position of a moving mirror by computing the spectral distribution and rate of particle production. With this, we can investigate the degree of asymmetry in the system by comparing the particle production by each of the two sides of the mirror. 

Given that the motion of the mirror and the varying of its material properties can be viewed as two simultaneous and distinct sources, interference effects will arise from the interaction of the two sources and modify the total spectrum and degree of asymmetry in the system \cite{silva2015interference}. This is not unexpected, as this emergent effect has been seen in similar DCE systems \cite{ji1998interference,lambrecht1998frequency,dalvit1999creation,alves2010quantum,silva2015interference}. In these systems the relationship between the relative oscillation frequency and phase difference of the distinct sources gives rise to constructive and
destructive interference in the spectral distribution of
created particles.

Highlighted in \cite{ji1998interference,silva2015interference}, the modified DCE spectrum, now accounting for the interference between the two sources, closely resembles the formula for the wave intensity of the double-slit interference experiment (or more generally, any monochromatic two source interference). Specifically, the authors in \cite{silva2015interference} observe an analogous formula relating the spectra of the distinct fluctuation source and the interference term. We will also obtain a corresponding formula, whereby we show that the asymmetric components of the different spectra (spectral differences) can also be related to each other in a similar manner.

This paper is organized as follows. In Sec. \ref{sec:scat}, we use a scattering approach \cite{jaekel1991casimir,jackel1992fluctuations,jaekel1997movement} to determine how the movement of the mirror and its time-varying material properties each modify the outgoing field as the mirror interacts with the vacuum. In Sec. \ref{sec:mixed}, we compute the full spectrum of created particles and the rate of particle creation, highlighting the contribution from an interference term that can contribute in a constructive and destructive way. In Sec. \ref{sec:diff}, we investigate the asymmetric contribution to the spectrum and comment on how the different contributions to the asymmetry are effected by the oscillation frequency of the mirror's time-dependent components. In Sec. \ref{sec:enhance}, we consider a stationary, asymmetric $\ddp$ mirror with two distinct sources modifying the mirror's properties, exploring the enhancement to the asymmetry of the system, and generalizing this to a system with an arbitrary number of distinct fluctuation sources. Final results are presented in Sec. \ref{sec:final}.

\section{\label{sec:scat}The Scattering Framework}

We start by considering a mirror, at rest, interacting with a real and massless scalar field in (1+1)D. Due to the presence of the mirror, fixed at $x=0$, we may decompose the field as
\begin{equation}\label{scalar field}
    \phi(t,x)=\Theta(x)\phi_+(t,x)+\Theta(-x)\phi_-(t,x),
\end{equation}
where $\Theta(x)$ is the Heaviside step-function and $\phi_+$ ($\phi_-$) is the field on the right (left) side of the mirror. Since both of $\phi_\pm$ obey the Klein--Gordon equation individually, they can be represented by the sum of two freely counterpropagating fields. In the frequency domain these are
\begin{equation}\label{phi p}
        \phi_+(t,x)=\int \frac{\mathrm{d}\w}{\sqrt{2\pi}}\qty[\phi_{\text{out}}(\w)e^{i\w x}+\psi_{\text{in}}(\w)e^{-i\w x}]e^{-i\w t}
\end{equation}
and
\begin{equation}\label{phi m}
        \phi_-(t,x)=\int \frac{\mathrm{d}\w}{\sqrt{2\pi}}\qty[\phi_{\text{in}}(\w)e^{i\w x}+\psi_{\text{out}}(\w)e^{-i\w x}]e^{-i\w t},
\end{equation}
where the amplitudes of the incoming and outgoing fields are labeled accordingly.

The incoming fields are unaffected by the mirror and take the standard form
\begin{equation}\label{waves in 1}
        \phi_{\text{in}}(\w)=(2\abs{\w})^{-1/2}\qty\big[\Theta(\w)a_L(\w)+\Theta(-\w)a_L^\dagger(-\w)]
\end{equation}
\noindent and
\begin{equation}\label{waves in 2}
        \psi_{\text{in}}(\w)=(2\abs{\w})^{-1/2}\qty\big[\Theta(\w)a_R(\w)+\Theta(-\w)a_R^\dagger(-\w)],
\end{equation}
where $a_j(\w)$ and $a_j^\dagger(\w)$ {$(j=L,R)$} are the annihilation and creation operators for the left ($L$) and right ($R$) sides of the mirror. These operators obey the commutation relation
\begin{equation}
[a_i(\w),a_j^\dagger(\w')]=\delta(\w-\w')\delta_{ij}, 
\end{equation}
where $\delta_{ij}$ is the Kronecker delta. 

The ingoing and outgoing fields are linearly related as
\begin{equation}\label{scat 0}
    \Phi_{\text{out}}(\w)=S(\w)\Phi_{\text{in}},
\end{equation}
where $S(\w)$ is the most general partially reflecting scattering matrix. Explicitly it is,
\begin{equation}
    S(\w)=\begin{pmatrix} s_+(\w) & r_+(\w) \\ r_-(\w) & s_-(\w) 
    \end{pmatrix}\, 
\end{equation}
where $r_\pm\left(\omega\right)$ and $s_\pm\left(\omega\right)$ are the reflection and transmission coefficients, respectively. These totally describe the effect of the mirror on the fields.
Here, we are making use of the vectorized shorthand
\begin{equation}
    \Phi_{\text{in}}(\w)=\begin{pmatrix}\phi_{\text{in}}(\w)\\\psi_{\text{in}}(\w)
    \end{pmatrix}~~~\text{and}~~~\Phi_{\text{out}}(\w)=\begin{pmatrix}\phi_{\text{out}}(\w)\\\psi_{\text{out}}(\w)
    \end{pmatrix}
\end{equation}
to represent ingoing and outgoing fields. In any situation where $\Phi(\omega)$ is used without a subscript, it can be assumed that the given relation holds for both ingoing and outgoing fields.

Up to this point, the properties of the mirror have remained general. Henceforth, we will consider the mirror interaction described by the asymmetric, partially reflected $\ddp$ mirror, whose potential is given as
\begin{equation}\label{potential}
    U(x)=\mu \delta(x)+\lambda \delta^\prime(x).
\end{equation}
Here, $\mu$ is 
related to the plasma frequency of the mirror and $\lambda$ is a dimensionless factor. The explicit form of the scattering matrix transmission and reflection components can be found to be \cite{silva2016dynamical}
\begin{equation}\label{gen coef1}
        r_{\pm}(\w)=\frac{-i\mu_0\pm2\w\lambda_0}{i\mu_0+\w(1+\lambda_0^2)}
\end{equation}
and
\begin{equation}\label{gen coef2}     
        s_{\pm}(\w)=\frac{\w(1-\lambda_0^2)}{i\mu_0+\w(1+\lambda_0^2)},
\end{equation}
where we introduce the notations $\mu_0$ and $\lambda_0$ to explicitly denote these as the zeroth-order terms. This distinction becomes
important as we start to include perturbative effects {below}. 

Let us first begin with the derivation of the $\ddp$ mirror undergoing mechanical oscillations about $x=0$. Scattering is still linear with
\begin{equation}\label{inert scat}
        \Phi'_{\text{out}}(\w)=S(\w)\Phi'_{\text{in}},
\end{equation}
in the co-moving frame (denoted by primes). In this frame the mirror is instantaneously at rest. The movement is assumed to be nonrelativistic ($\abs{\Dot{q}(t)}\ll1$) and limited by a small amplitude, such that the mirror's position becomes
\begin{equation}
q(t) = \epsilon g(t),   
\end{equation}
with $|g(t)| \leq 1$ and $\epsilon\ll1$. To solve this in the laboratory frame, we use the relation
\begin{equation}\label{frame relation}
    \Phi'(t',0)=\Phi(t,\epsilon g(t))=[1-\epsilon g(t)\eta\partial_t]\Phi(t,0)+\mathcal{O}(\epsilon^2),
\end{equation}
where $\eta=\text{diag}(1,-1)$. Taking advantage of the fact that $\mathrm{d}t'=\mathrm{d}t$ to order $\epsilon^2$, \eqref{frame relation} can be rewritten as 
\begin{equation}
\label{temp transf}
\Phi'(t,0)=[1-\epsilon g(t)\eta\partial_t]\Phi(t,0). 
\end{equation}
We find that applying this transform to \eqref{inert scat} in the frequency domain yields
\begin{equation}
    \Phi_{\text{out}}(\w)=S_0(\w)\Phi_{\text{in}}(\w)+\int\frac{\mathrm{d}\w'}{2\pi}\delta S_q(\w,\w')\Phi_{\text{in}}(\w'),
\end{equation}
where we suppress the evaluation of $x=0$ in $\Phi(\w,0)$ going forward. The first order \textit{S}-matrix, $\delta S_q(\w,\w')$, takes the form
\begin{equation}
    \delta S_q(\w,\w')=i \epsilon\w' \mathcal{G}(\w-\w')[S_0(\w)\eta-\eta S_0(\w')],
\end{equation}
where $\mathcal{G}(\w)$ is the Fourier transform of $g(t)$, and $S_0$ is the zeroth-order scattering matrix found from {Eqs.} 
\eqref{gen coef1} and \eqref{gen coef2}. This is in agreement with \cite{silva2016dynamical}.

Now, let us solve the ADCE for the $\ddp$ mirror with time-dependent $\mu(t)$. For this analysis, we assume that the mirror is held at rest. Here we require fluctuations in $\mu(t)$ take the form of small oscillations about a fixed value $\mu_0$. Specifically, 
\begin{equation}
    \mu(t)=\mu_0\qty[1+\epsilon f(t)],
\end{equation}
where $\mu_0\geq1$ is a constant and $f(t)$ is an arbitrary function such that $\abs{f(t)}\leq1$, with $\epsilon\ll1$. 

To find the modified outgoing field, we apply the field equation of the system, determined by the potential \eqref{potential}, to Eqs. \eqref{phi p} and \eqref{phi m}. From here, the matching conditions can be solved to the first order, where the final form of $\Phi_{\text{out}}(\w)=S(\w)\Phi_{\text{in}}$ becomes
\begin{equation}\label{Phi u}
    \Phi_{\text{out}}(\w)=S_0(\w)\Phi_{\text{in}}(\w)+\int\frac{\mathrm{d}\w'}{2\pi}\delta S_\mu(\w,\w')\Phi_{\text{in}}(\w').
\end{equation}
The asymmetric correction that originates from the introduction of $f(t)$ takes the form 
\begin{equation}
    \delta S_\mu(\w,\w')=\epsilon\alpha_\mu(\w,\w')\mathbb{S}_\mu(\w'),
\end{equation}
where 
\begin{equation}\label{mu a S1}
  \alpha_\mu(\w,\w')=-\frac{i\mu_0\mathcal{F}(\w-\w')}{i\mu_0+\w(1+\lambda_0^2)} 
\end{equation}
and
\begin{equation}\label{mu a S2}
  \mathbb{S}_\mu(\w')=\begin{pmatrix} s_+(\w') & 1+r_+(\w') \\ 1+r_-(\w') &  s_-(\w')
    \end{pmatrix}.
\end{equation}
Here, $\mathcal{F}(\w)$ is the Fourier transform of $f(t)$. This is in agreement with \cite{silva2020motion}.

To summarize, we can now write the final form of the field for the $\ddp$ with the two first order perturbations. One that arises from fluctuations in position and another that arises due to fluctuations of a material property. Explicitly,
\begin{eqnarray}
\Phi_{\text{out}}(\w)=&&~ S_0(\w)\Phi_{\text{in}}(\w)+\int\frac{\mathrm{d}\w'}{2\pi}\delta S_\mu(\w,\w')\Phi_{\text{in}}(\w')\nonumber\\&&+\int\frac{\mathrm{d}\w'}{2\pi}\delta S_q(\w,\w')\Phi_{\text{in}}(\w'), 
\end{eqnarray}
with $\delta S(\w,\w')=\epsilon\alpha(\w,\w')\mathbb{S}(\w,\w')$, where 
\begin{equation}\label{q a S}
  \alpha_q(\w,\w')=i\w'\mathcal{G}(\w-\w'),~~~  \mathbb{S}_q(\w,\w')=S_0(\w)\eta-\eta S_0(\w')
\end{equation}
for the fluctuation in position and
\begin{equation}\label{mu a S}
  \alpha_\mu(\w,\w')=-\frac{i\mu_0\mathcal{F}(\w-\w')}{i\mu_0+\w(1+\lambda_0^2)},~~~  \mathbb{S}_\mu(\w')=J_2+S_0(\w')
\end{equation} for fluctuation in material properties. $J_2$ is the $2\times2$ column-reversed identity matrix. These results of the perturbation of the field due to the two separate fluctuation sources will be used in the next section to compute the full spectrum of particles for this mixed $\ddp$ system.

\section[Interference for a moving \texorpdfstring{$\ddp$}{TEXT} mirror with time-dependent \texorpdfstring{$\mu(t)$}{TEXT}]{Interference for a moving \texorpdfstring{$\ddp$}{TEXT} mirror with time-dependent \texorpdfstring{$\mu(t)$}{TEXT}}\label{sec:mixed}

The spectral distribution of created particles is given by \cite{lambrecht1996motion}
\begin{equation}\label{spec}
    N(\w)=2\w\Tr[\!\matrixel{0_{\text{in}}}{\Phi_{\text{out}}(-\w)\Phi_{\text{out}}^{\mathrm{T}}(\w)}{0_{\text{in}}}].
\end{equation}
Assuming the incoming waves are vacuum states, and making use of the following formula,
\begin{equation}
    \matrixel{0_{\text{in}}}{\Phi_{\text{in}}(\w)\Phi_{\text{in}}^{\mathrm{T}}(\w')}{0_{\text{in}}}=\frac{\pi}{\w}\delta(\w+\w')\Theta(\w),
\end{equation}
the total spectral contribution becomes 
\begin{equation}
     N(\w)=\frac{1}{2\pi}\int_0^{\infty}\frac{\mathrm{d}\w'}{2\pi}\frac{\w}{\w'}\Tr[\delta S(\w,-\w')\delta S^{\dagger}(\w,-\w')],
\end{equation}
where $\delta S=\delta S_\mu+\delta S_q$. 

The spectrum can be decomposed into several distinct contributions. Two result from the initial fluctuations in the mirror's position and properties, and a third results from the interference between the two independent sources of particle creation. Explicitly, 
\begin{equation}\label{full creation}
    N_\pm(\w)=N_{q\pm}(\w)+N_{\mu\pm}(\w)+N_{\mathrm{int}\pm}(\w),
\end{equation}
where we decompose the spectral distribution such that $N(\w)=N_+(\w)+N_-(\w)$, where $N_+(\w)$ ($N_-(\w)$) is the spectrum produced on the right (left) half of the mirror. The first independent contribution in Eq. \eqref{full creation} is due to the movement of the mirror, given by 
\begin{eqnarray}\label{Nq base}
    N_{q\pm}(\w)=&&\frac{\epsilon^2}{\pi}\int_0^\infty\frac{\mathrm{d}\w'}{2\pi}\w\w'\abs{\mathcal{G}(\w+\w')}^2\\&&\times\Re\qty[\frac{ i\mu_0 (1\mp\lambda_0)^2 (\w +\w')+8 \lambda_0^2 \w\w'-2 \mu_0^2 }{(i\mu_0 + \w(1 + \lambda_0^2))[i\mu_0 + \w'(1 + \lambda_0^2)]}],\nonumber
\end{eqnarray}
and the second term originates from the time-dependence of $\mu(t)$ of the material property of the mirror, given by
\begin{eqnarray}\label{Nu base}
    N_{\mu\pm}(\w)=~&&\frac{\epsilon^2\mu_0^2}{\pi}(1\pm\lambda_0)^2(1+\lambda_0^2)\\&&\times\int_0^\infty\frac{\mathrm{d}\w'}{2\pi}\Upsilon(\w)\Upsilon(\w')\abs{\mathcal{F}(\w+\w')}^2,\nonumber
\end{eqnarray}
where $\Upsilon(\w)=\w/\qty[\mu_0^2 + \w^2(1 + \lambda_0^2)^2]$. These are in agreement with \cite{silva2016dynamical} and \cite{silva2020motion}, respectively. The last term describes the interference effects in the system, taking the form
\begin{eqnarray}\label{Ni base}
        &&N_{\mathrm{int}\pm}(\w)=\frac{\epsilon^2\mu_0}{\pi}(1\pm\lambda_0)^2\int_0^\infty\frac{\mathrm{d}\w'}{2\pi}\Upsilon(\w)\Upsilon(\w')\\&&\times2\qty[\pm\mu_0^2-2\lambda_0(1+\lambda_0^2) \w\w']\Re\qty[\mathcal{G}(\w+\w')\mathcal{F}^*(\w+\w')].\nonumber
\end{eqnarray}
This term can exhibit both constructive and destructive interference, which arises from the fact that the motion of the mirror and its time-dependent properties act as two distinct sources of particle creation \cite{silva2015interference}.

To further investigate the particle creation due to the interference term, we consider the typical functions used to describe the motion of the mirror, 
\begin{equation}\label{f(t)}
    f(t)=\cos(\w_1t)\exp(-\abs{t}/\tau)
\end{equation}
and
\begin{equation}\label{g(t)}
    g(t)=\cos(\w_2t+\phi)\exp(-\abs{t}/\tau)
\end{equation}
where the frequencies of oscillation are $\w_1$ and $\w_2$, with $\tau$ being the effective oscillation time of the system and $\phi$ is a constant phase. Only the monochromatic limit is considered, with $\w_1\tau\gg1$ and $\w_2\tau\gg1$. In this limit the system undergoes an effectively spatially symmetric motion about its starting position.
In the monochromatic limit \cite{silva2015interference, silva2011simple, mintz2006particle} we see
\begin{equation}\label{mono}
    \lim_{\tau \to \infty} \frac{\abs{\mathcal{F}(\w)}^2}{\tau}=\frac{\pi}{2}\qty[\delta(\w-\w_0)+\delta(\w+\w_0)],
\end{equation}
where the same relation holds for $\mathcal{G}(\w)$. Using Eq. \eqref{mono}, we see that 
\begin{eqnarray}\label{Nq}
    &&\frac{N_{q\pm}(\w)}{\tau}=\frac{\epsilon^2}{4\pi}\w(\w_1-\w)\Theta(\w_1-\w)\\&&~~~\times\Re\qty[\frac{ i\mu_0 (1\mp\lambda_0)^2 \w_1+8 \lambda_0^2 \w(\w_1-\w)-2 \mu_0^2 }{(i\mu_0 + \w(1 + \lambda_0^2))[i\mu_0 + (\w_1-\w)(1 + \lambda_0^2)]}],\nonumber
\end{eqnarray}
which is in agreement with \cite{silva2016dynamical}, and
\begin{equation}\label{Nu}
    \frac{N_{\mu\pm}(\w)}{\tau}=\frac{\epsilon^2\mu_0^2}{4\pi}(1\pm\lambda_0)^2(1+\lambda_0^2)\Upsilon(\w)\Upsilon(\w_2-\w)\Theta(\w_2-\w),
\end{equation}
which is consistent with \cite{silva2020motion}. 

Due to the different oscillation frequencies of the independent sources, $\w_1$ and $\w_2$, the calculation of the interference term needs to be carefully considered. From \cite{silva2015interference}, we see that the interference term $N_{\mathrm{int}\pm}(\w)/\tau$ vanishes when $\w_1\neq\w_2$, due to the term involving $\Re[\mathcal{G}(\w)\mathcal{F}^*(\w)]/\tau$ in  Eq. \eqref{Ni base}. However, when $\w_1=\w_2=\w_0$ the interference term becomes
\begin{eqnarray}\label{Nint}
    \frac{N_{\mathrm{int}\pm}(\w)}{\tau}=&&\frac{\epsilon^2\mu_0}{2\pi}(1\pm\lambda_0)^2\qty[\pm\mu_0^2-2\lambda_0(1+\lambda_0^2)\w(\w_0-\w) ]\nonumber\\&&\times\Upsilon(\w)\Upsilon(\w_0-\w)\Theta(\w_0-\w)\cos\phi.
\end{eqnarray}

\begin{figure}
\includegraphics[width=\columnwidth]{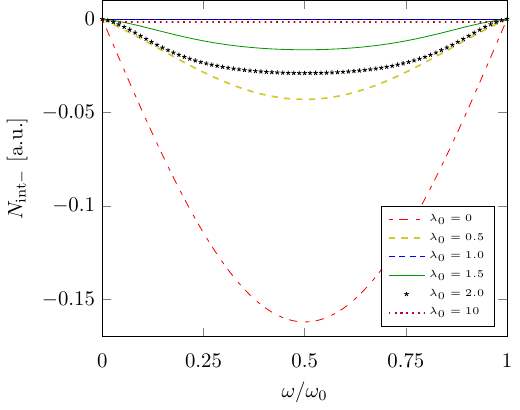}
\caption{\label{figure:spectrum int L}The spectral distribution due to the interference term $N_{\mathrm{int}-}(\w)/(2\epsilon^2\tau\pi^{-1})$ on the left half of the mirror ($x<0$) for some values of $\lambda_0$, with $\uz=1$ and $\phi=0$.}
\end{figure}

As previously noted, this spectrum exhibits both constructive and destructive interference. The exact manner of this interference will depend on its material properties and degree of asymmetry. In addition, the phase $\phi$, from the oscillation definition in \eqref{g(t)}, will alter the sign on the interference term. From Eq.~\eqref{Nint}, it is easy to show that the spectrum on the left side of the mirror, seen in Fig. \ref{figure:spectrum int L}, will always have the same sign for any $\w$, thus we will only have destructive interference if $0\leq\phi<\pi/2$ and constructive if $\pi/2\leq\phi<\pi$.

\begin{figure}
\includegraphics[width=\columnwidth]{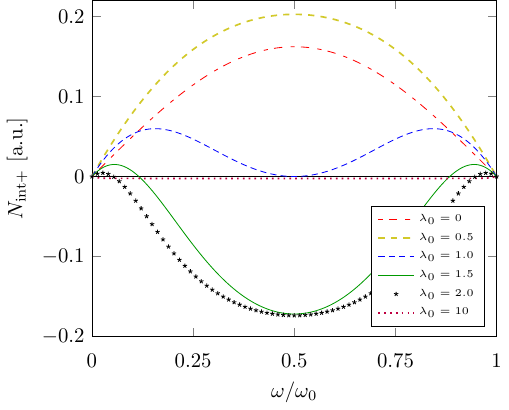}
\caption{The spectral distribution due to the interference term $N_{\mathrm{int}+}(\w)/(2\epsilon^2\tau\pi^{-1})$ on the right half of the mirror ($x>0$) for some values of $\lambda_0$, with $\uz=1$ and $\phi=0$.}
\label{figure:spectrum int R}
\end{figure}

The spectrum of the right half of the mirror, seen in Fig. \ref{figure:spectrum int R}, is more complicated, where we now see the only region whose sign remains constant for any $\w$ occurs when $\lambda_0(1+\lambda_0^2)\w_0^2 < 2\mu_0^2$. Again we see the interference effect change for different phases, except now we have constructive interference if $0\leq\phi<\pi/2$ and destructive if $\pi/2\leq\phi<\pi$. However, when $\lambda_0(1+\lambda_0^2)\w_0^2 > 2\mu_0^2$, we can now solve for the two real roots (symmetrical with respect to $\w_0/2$) with the following equation, 
\begin{equation}\label{int w}
    2\w_\pm=\w_0\pm\sqrt{\w_0^2-\frac{2\mu_0^2}{\lambda_0(1+\lambda_0^2)}}.
\end{equation}
Note that $\w_\pm$ here represents only the positive and negative roots for the right side of the mirror and does not denote a frequency with respect to the two halves. For $0\leq\phi<\pi/2$, the interference is constructive for $\w<\w_-$ and $\w>\w_+$ and destructive for $w_-<\w<\w_+$, while the opposite occurs if $\pi/2\leq\phi<\pi$. In the limiting case of $\lambda_0(1+\lambda_0^2)\w_0^2 \gg 2\mu_0^2$ the sign of the interference does not change, with $\w_-\xrightarrow[]{}0$ and $\w_+\xrightarrow[]{}\w_0$. In this regime the interference of the right side of the mirror will be  destructive if $0\leq\phi<\pi/2$ and constructive if $\pi/2\leq\phi<\pi$.

The spectral distribution for each of the separate contributions in Eqs. \eqref{Nq}, \eqref{Nu}, and \eqref{Nint} is limited by the oscillation frequency, with no particles being produced with frequency greater than $\w_0$ when the interference term is involved. The spectrums are symmetric with respect to $\w=\w_0/2$, as it is invariant under the change $\w\xrightarrow{}\w_0-\w$. This is a consequence of the fact that particles are produced in pairs, where one is produced with a frequency $\w$ and the other with frequency $\w_0-\w$ \cite{mintz2006particle,silva2015interference,rego2013inhibition,silva2011simple,silva2016dynamical}.

Using the spectral distribution for the interference term in Eq. \eqref{Nint}, we are able to calculate how the rate of particle production is affected by the addition of the interference effect. The total number of created particles is given by
\begin{equation}
    \mathcal{N}=\int_0^\infty\frac{\mathrm{d}\w}{2\pi}N(\w),
\end{equation}
whereby the particle creation rate is given by $\mathcal{N}/\tau$, with $\mathcal{N}=\mathcal{N}_q+\mathcal{N}_\mu+\mathcal{N}_\mathrm{int}$. The independent contribution due to the motion of the mirror can be expressed as 
\begin{align}
    \mathcal{N}_q&=(\epsilon^2\tau\w_0^3/\pi)\mathcal{F}(\xi)
\end{align}
where $\xi=(1+\lz^2)\w_0/\uz$  and
\begin{equation}
    \mathcal{F}(\xi)=\frac{\mathcal{A}(\xi)+\mathcal{B}(\xi)\ln[\xi^2+1]+\mathcal{C}(\xi)\arctan\xi}{6\xi^3(1+\lz^2)^2[\xi^2+4]}.
\end{equation}
The explicit form of $\mathcal{A}, ~\mathcal{B}, ~\mathrm{and}~ \mathcal{C}$ are lengthy and, for brevity, we refer to \cite{silva2016dynamical} for the complete forms of these expressions. We will, however, present the complete form of the creation rate solely originating from the changing material properties of the mirror, as its explicit form has not been published before. We find that 
\begin{align}
    \mathcal{N}_\mu&=(\epsilon^2\tau\w_0/\pi)\mathcal{G}(\xi),
\end{align}
where
\begin{equation}
    \mathcal{G}(\xi)=\frac{(\xi^2+2)\ln[1+\xi^2]-2\xi\arctan\xi}{2\xi^2(\xi^2+4)}.
\end{equation}
The particle creation rate due to the time-varying boundary conditions takes a near identical form of the creation rate of a mirror described by the time-varying Robin boundary condition. These two cases can be related to each other under the condition $\lambda=1$, with $\gamma_0=2/\uz$ being the Robin parameter \cite{silva2016dynamical}. Under these conditions, the creation rate for the $\ddp$ mirror with fluctuating properties can be related to that of the stationary, time-dependent Robin mirror via $\mathcal{N}_\mu=\w_0^2\mathcal{N}_\gamma$.

The interference term \eqref{Nint} takes the form
\begin{align}\label{creation rate int}
    \mathcal{N}_\mathrm{int}&=(2\epsilon^2\tau\lz\w_0^3\cos{\phi}/\uz\pi)\mathcal{I}(\xi),
\end{align}
where 
\begin{equation}
    \mathcal{I}(\xi)=\frac{\xi(\ln[1+\xi^2]-4-\xi^2)+2(2+\xi^2)\arctan\xi}{\xi^3(\xi^2+4)}.
\end{equation}
We can compare this result to that of the interference in the creation rate of the moving, time-dependent Robin mirror \cite{silva2015interference}. Just as before, with $\lambda=1$ and $\gamma_0=2/\uz$, we find that the interaction creation rates are nearly identical, up to an overall factor of $-2\w_0$. From Eq. \eqref{creation rate int}, it is evident that the nature of the interference will depend on the value of $\xi$ and the phase angle $\phi$. When $0\leq\phi<\pi/2$, the interference is constructive for $0<\xi<2.23$ and destructive for $\xi>2.23$. The reverse occurs for when $\pi/2<\phi\leq2\pi$. The value of $\xi\approx2.23$ is consistent when compared to the inflection point of the moving, time-dependent Robin mirror. This is not unexpected, as this value corresponds to a general inhibition of particle creation for the Robin mirror \cite{silva2016dynamical,silva2015interference}. For this specific value of $\xi$, $\mathcal{G}(\xi)=0$. The interference creation rate will completely vanish when $\phi=\pi/2$.

\section{\label{sec:diff}Asymmetric Particle 
Production}

In order to quantify and understand the asymmetry present in the mixed $\ddp$ system, we will investigate the difference in particle spectrum  by the two sides of the mirror. This quantity, $\Delta N(\w)=N_-(\w)-N_+(\w)$, is a useful tool in quantifying the ADCE spectrum. For the simplicity, and the sake of a comparison, we will use $\w_1=\w_2=\w_0$. Just as we presented in \eqref{full creation} we can present the total difference between the two sides as
\begin{equation}\label{DN}
    \Delta N(\w)=\Delta N_q(\w)+\Delta N_\mu(\w)+\Delta N_{\mathrm{int}}(\w).
\end{equation}
The spectral difference for the particle spectrum generated by the motion of the mirror
\begin{equation}\label{DN moving dd}
    \begin{split}
        \frac{\Delta N_q(\w)}{\tau}&=\frac{\epsilon^2}{\pi}\w_0^2\mu_0^2\lambda_0(1+\lambda_0^2)\Upsilon(\w)\Upsilon(\w_0-\w)\Theta(\w_0-\w)
    \end{split}
\end{equation}
and from its changing properties
\begin{equation}\label{DN properties}
    \begin{split}
        \frac{\Delta N_\mu(\w)}{\tau}&=-\frac{\epsilon^2}{\pi}\mu_0^2\lambda_0(1+\lambda_0^2)\Upsilon(\w)\Upsilon(\w_0-\w)\Theta(\w_0-\w)\\
    \end{split}
\end{equation}
which leads to the relationship
\begin{equation}\label{DN resonence}
    \Delta N_q(\w)=-\w_0^2\Delta N_\mu(\w).
\end{equation}

From Eq. \eqref{DN resonence}, we see clear resonance behavior exhibited by the mixed $\ddp$ system. The system is in total resonance when $\w_0=1$, with $\Delta N_q(\w)=-\Delta N_\mu(\w)$, and the only contribution to the total difference between the two sides is the spectral difference of the interference $\Delta N_{\mathrm{int}}(\w)$. Note that, even though the total spectral difference can appear to be negative, the particle production by each side will always be positive. A negative (positive) spectral difference indicates that the right (left) half of the mirror produces the greater number of particles. From the form of resonance relationship in Eq. \eqref{DN resonence}, we see that the two distinct sources of particle production, $N_q$ and $N_\mu$, oppose each other. The asymmetry present in the production of particles for these two fluctuation sources oppose each other, each reducing the other's contribution to the total imbalance of the system.

From Eq. \eqref{DN resonence}, it is clear that in the low driving frequency regime, $\w_0\ll 1$, the asymmetry in particle production will be dominated by the contribution from the time-varying properties of the mirror. The reverse is true for the high frequency regime, where the dominant asymmetric contribution comes from the physical oscillation of the mirror when $\w_0\gg 1$. In the low oscillation frequency regime, the dominant contribution to the total spectral difference term is now
\begin{equation}\label{low freq}
    \Delta N_\mu(\w)\approx\frac{\lz(1+\lz^2)\uz^2\w^2}{\pi[\uz^2+\w^2(1+\lz^2)^2]^2}  \quad\quad\quad\w_0\ll1
\end{equation}
In the high oscillation frequency regime, the dominant contribution to the total spectral difference term is now
\begin{equation}\label{high freq}
    \Delta N_q(\w)\approx\frac{\lz\uz^2\w(\w_0+\w)}{\pi(1+\lz^2)[\uz^2+\w^2(1+\lz^2)^2]}  \quad\quad\quad\w_0\gg1
\end{equation}

\begin{figure}
\includegraphics[width=\columnwidth]{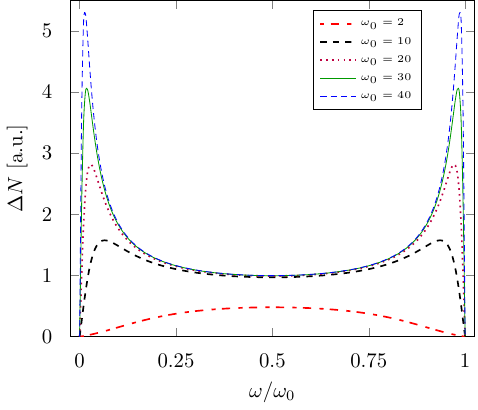}
\caption{The total spectral difference $\Delta N(\w)/(\tau\epsilon^2\pi^{-1})$ for different values of $\w_0$, with $\uz=\lz=1$ and $\phi=0$.}
\label{figure:spectrum tot DIFF}
\end{figure}

We can see the effects the oscillation frequency has on the total spectral difference in Fig. \ref{figure:spectrum tot DIFF}. The higher the oscillation frequency $\w_0$ becomes, the more pronounced the two peak behaviour in $\Delta N$ becomes. These two peaks emerge near $\w\approx0$ and $\w_0$. When the phase shift $\phi$ in the interference term is anti-aligned ($\phi=\pi$), the two peak behavior will dominate the total spectral difference as the spectrum near $\w=\w_0/2$ is suppressed, with $\Delta N(\w_0/2)=0$.

\begin{figure}
\includegraphics[width=\columnwidth]{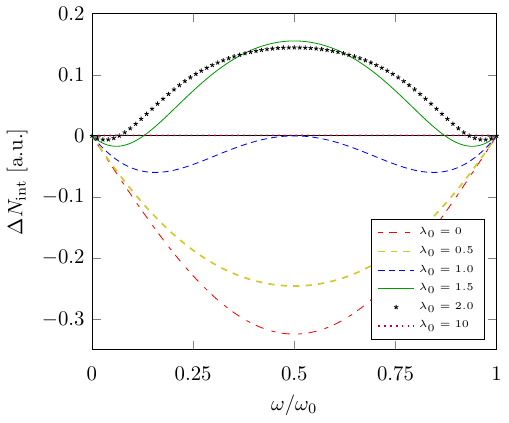}
\caption{The spectral difference from the interference term $\Delta N_{\mathrm{int}}(\w)/(2\epsilon^2\tau\pi^{-1})$ for some values of $\lambda_0$, with $\uz=1$ and $\phi=0$.}
\label{figure:spectrum int DIFF}
\end{figure}

The spectral difference that comes from the interference effect between these two particle sources is 
\begin{eqnarray}\label{DN int}
    \frac{\Delta N_{\mathrm{int}}}{\tau}=-\frac{\epsilon^2}{\pi}\mu_0&&(1+\lambda_0^2)\qty[\mu_0^2-4\lambda_0^2\w(\w_0-\w)]\\&&\times\Upsilon(\w)\Upsilon(\w_0-\w)\Theta(\w_0-\w)\cos\phi.\nonumber
\end{eqnarray}
We can relate these differences in the following manner,
\begin{equation}\label{double slit form}
    \abs{\Delta N_{\mathrm{int}}(\w)}=I(\w)2\sqrt{\abs{\Delta N_q(\w)\Delta N_\mu(\w)}}\cos\phi
\end{equation}
where the interference term $I(\w)$ takes the form
\begin{equation}
    I(\w)=(2\lambda_0\mu_0\w_0)^{-1}\qty[\mu_0^2-4\lambda_0^2\w(\w_0-\w)].
\end{equation}
The difference in right and left half constructive and destructive interference will again be dependent on the region and form of the system. Now, the only region whose sign remains constant for any $\w$ occurs when $\lambda_0\w_0 < \mu_0$, where we have destructive interference if $0\leq\phi<\pi/2$ and constructive if $\pi/2\leq\phi<\pi$. For the case when $\lambda_0\w_0 > \mu_0$, we can again solve for the two real roots using the following equation, 
\begin{equation}\label{DNi roots}
    2\w_\pm=\w_0\pm\sqrt{\w_0^2-\mu_0^2/\lambda_0^2}.
\end{equation}
For $0\leq\phi<\pi/2$ the interference is destructive for $\w<\w_-$ and $\w>\w_+$ and constructive for $\w_-<\w<\w_+$, while the opposite occurs if $\pi/2\leq\phi<\pi$. In the limiting case of $\lambda_0\w_0 \gg \mu_0$, the interference will be  constructive if $0\leq\phi<\pi/2$ and destructive if $\pi/2\leq\phi<\pi$. These characteristics can been seen in Fig. \ref{figure:spectrum int DIFF}.

For $0\leq\phi<\pi/2$, the interference term will produce more particles, while the opposite occurs if $\pi/2\leq\phi<\pi$. In the limiting case of $\lambda_0(1+\lambda_0^2)\w_0^2 \gg 2\mu_0^2$ the sign of the interference effect becomes effectively constant for any $\w$, with $\w_-\xrightarrow[]{}0$ and $\w_+\xrightarrow[]{}\w_0$. In this regime the interference of the right side of the mirror will be  destructive if $0\leq\phi<\pi/2$ and constructive if $\pi/2\leq\phi<\pi$.

It is convenient to write the total spectral difference in the following form,
\begin{equation}\label{DN tot}
    \Delta N(\w) = \qty[1+2\w_0 I(\w)\cos\phi-\w_0^2]\Delta N_\mu(\w).
\end{equation}
We can now clearly see how the different values of the frequency and phase shift will effect the magnitude and direction of the asymmetry. For positive values of $\lambda_0$, the spectral difference for the time-varying properties contribution in Eq. \eqref{DN properties} is always going to be negative; the right side of the mirror will always produce more particles than the left side. Thus, the total spectral difference \eqref{DN tot} will also be negative when $\w_0^2-2\w_0I(\w)\cos{\phi}>1$ and will be positive for the reverse. The asymmetry will vanish ($\Delta N(\w)=0$) when $\w_0^2-2\w_0I(\w)\cos{\phi}=1$.

\section{Asymmetric Enhancement}\label{sec:enhance}
Amplification of the DCE particle production is an important consideration when designing experimental tests of this phenomenon. Just as Moore pointed out, the particle production in DCE systems is limited \cite{moore1970quantum}, so any means by which the spectrum can be enhanced will allow for easier and better experimental measurements. Within the context of the ADCE, enhancement of the spectral output will couple with the asymmetry of the system to produce even greater imbalances between the two sides of the mirror. This leads to an increase in the magnitude of the unbalanced radiation pressure, resulting in a larger net force which induces motion on an otherwise stationary mirror.

For the $\ddp$ mirror setup we have considered up to this point, it is evident from Eq. \eqref{DN resonence} that the individual particle spectrum of the two distinct fluctuation sources, the time-varying properties and the movement of the mirror, work in opposition to each other. However, it is possible to construct a system whose creation terms work in conjunction with each other and whose interference pattern allows the asymmetry of the system to be further enhanced (or reduced). 

Now, we will modify our asymmetric setup in a similar manner as the construction of a mixed SQUID system with two independent sources of magnetic flux \cite{silva2015interference}, where both sources driving harmonic variations in the Josephson energy of the SQUID present different phases and frequencies. Here, we consider a stationary $\ddp$ mirror with two distinct, unspecified fluctuation sources that each modify the material properties in such a way that the new time-dependent $\mu(t)$ now takes the form
\begin{equation}\label{mu 2 source}
    \mu(t)\approx\mu_0\qty[1+\epsilon_1f_1(t)+\epsilon_2f_2(t)],
\end{equation}
where 
\begin{align}
    f_1(t)&=\cos(\w_1t)e^{-\abs{t}/\tau},\\[10pt]
    f_2(t)&=\cos(\w_2t+\phi)e^{-\abs{t}/\tau}.
\end{align}

Just as before, when $\w_1\neq\w_2$, the interference term vanishes and the total spectra for the two sides of the mirror, each taking the form of \eqref{Nu}, is now only the sum of the two independent particles spectrum:
\begin{equation}
    N_{\mu\pm}(\w)=N_{\mu\pm}^{(1)}(\w)+N_{\mu\pm}^{(2)}(\w)
\end{equation}
where
\begin{eqnarray}
    N_{\mu\pm}^{(1)}(\w)=\frac{\epsilon_1^2\tau}{4\pi}\mu_0^2(1\pm\lambda_0)^2(1+\lambda_0^2)&&\Upsilon(\w)\Upsilon(\w_1-\w)\label{enhance 1}\nonumber\\&&\times\Theta(\w_1-\w),\\[10pt]
    N_{\mu\pm}^{(2)}(\w)=\frac{\epsilon_2^2\tau}{4\pi}\mu_0^2(1\pm\lambda_0)^2(1+\lambda_0^2)&&\Upsilon(\w)\Upsilon(\w_2-\w)\label{enhance 2}\nonumber\\&&\times\Theta(\w_2-\w).
\end{eqnarray}

It is straightforward to calculate the total spectrum when $\w_1=\w_2=\w_0$, which now includes the contribution from the interference term. Since we are interested in amplifying the asymmetry of the system, we will focus on analyzing the enhancement of the spectral difference. The greater the spectral difference, the greater the asymmetry of the system, which leads to an increase in non-vanishing net force on the system due to the further imbalance in the radiation pressure from the production of real particles. The total spectral difference when the frequencies are in resonance, $\w_1=\w_2=\w_0$, is now
\begin{eqnarray}\label{enhance tot}
    \Delta N_\mu(\w)=\Delta N_\mu^{(1)}&&(\w)+\Delta N_\mu^{(2)}(\w)\\&&+2\sqrt{\Delta N_\mu^{(1)}(\w)\Delta N_\mu^{(2)}(\w)}\cos\phi\nonumber
\end{eqnarray}
Using Eqs. \eqref{enhance 1} and \eqref{enhance 2} in \eqref{enhance tot}, one can show that 
\begin{equation}\label{enhance new}
    \Delta N_\mu(\w)=-\frac{\epsilon(\phi)^2\tau}{\pi}\mu_0^2\lambda_0(1+\lambda_0^2)\Upsilon(\w)\Upsilon(\w_0-\w)\Theta(\w_0-\w),
\end{equation}
where
\begin{equation}\label{enhance eps}
    \epsilon(\phi)^2=\epsilon_1^2+\epsilon_2^2+2\epsilon_1\epsilon_2\cos{\phi},
\end{equation}
which is in agreement with \cite{silva2015interference}. The interference term in Eq. \eqref{enhance new} leads to either an increase or decrease in the magnitude of the asymmetry of the system. A maximum enhancement (reduction) of the asymmetry will occur when the relative phase between the two fluctuation sources are aligned (anti-aligned) when $\phi=0$ ($\phi=\pi$). In fact, for the case of two sources with $\epsilon_1=\epsilon_2$, the spectral difference completely vanishes in the anti-aligned case ($\epsilon(\phi)=0$). This results in a purely symmetric production of particles as this leads to the total spectral difference vanishing. When the two sources are out of phase ($\phi=\pi/2$), the resulting spectral difference reduces to only the contribution from the independent source terms, as expected.

To further maximize the enhancement of the asymmetry, it is natural to investigate the ramifications of adding an arbitrary number of distinct fluctuation sources. In this model, we will further modify Eq. \eqref{mu 2 source} in the following manner
\begin{equation}\label{full enhance}
    \mu(t)\approx\mu_0\qty[1+\sum_{i=1}^N\epsilon_if_i(t)],
\end{equation}
where 
\begin{equation}
    f_i(t)=\cos(\w_it+\phi_i)e^{-\abs{t}/\tau},
\end{equation}
and $\phi_1=0$, with $N$ being the number of distinct sources. Using our new definition of $\mu(t)$ for a generic number of fluctuation sources, we can find the expression for the new total spectral difference. It follows that $\Delta N(\w)$ will be of the same form as Eq. \eqref{enhance new}, except $\epsilon(\phi)^2$ becomes 
\begin{equation}\label{enhance eps gen}
    \epsilon(\phi)^2=\sum_{ i=1}^N\epsilon_i^2+\sum_{i\neq j}^N\epsilon_i\epsilon_j\cos{(\phi_j-\phi_i)}.
\end{equation}
As one would expect, to obtain the maximally asymmetric enhancement the relative phase shifts of the independent sources need to all be in-phase with one another. That is, $\phi_i=\phi_j$ for all $i$ and $j$. With this in mind, we will set all $\phi_i=\phi_j=0$. Additionally, we will normalize the magnitude of the different fluctuation sources such that $\epsilon_i=1$. In this maximally enhanced limit, we see that 
\begin{equation}\label{full enhanc}
    \epsilon(\phi)^2=(N\epsilon)^2.
\end{equation}
This shows that the ADCE exhibits a clear sign of two source monochromatic interference. When sources are totally coherent ($\Delta\phi=0$) contributions will add purely as amplitudes ($\epsilon=\sum\epsilon_i$) and when sources are totally incoherent ($\Delta\phi=\pi/2$) contributions add purely as intensities ($\epsilon^2=\sum\epsilon_i^2$). From Eq. \eqref{full enhanc}, we find that the total asymmetry of the system will increase by a factor of $N^2$ due to the additional independent sources acting on the stationary $\ddp$ mirror. 

\section{Final Remarks}\label{sec:final}
We investigated the interference effects that arise from the presence of multiple independent sources of particle creation in an ADCE system. This was modeled by a partially reflecting moving mirror simulated with a $\ddp$ point-like mirror interacting with a real massless scalar field in (1+1)D. Specifically, one of our models involves the interaction of a moving $\ddp$ mirror with time-dependent material properties $\mu(t)$. The other model explores the interaction between two independent sources of material property fluctuations of $\mu(t)$ for a stationary mirror. This analysis is expanded to account for an arbitrary number of independent field perturbation sources and its enhancement of the asymmetric spectrum produced. 

For the moving time-dependent $\ddp$ model, we find the spectral contribution from the motion \eqref{Nq base} (in agreement with \cite{silva2016dynamical}) and the fluctuating properties \eqref{Nu base} (in agreement with \cite{silva2020motion}), along with an interference term \eqref{Ni base} that arises from the interaction between the two distinct sources of particle creation. Using the typical functions to describe the time fluctuations \eqref{f(t)} and \eqref{g(t)} (which reduce to \eqref{mono} in the monochromatic limit), we see (as in \cite{silva2015interference}) that the interference term vanishes when the two perturbation sources are driven by two different oscillation frequencies and the system reduces to only the contribution from the independent sources in \eqref{Nq} and \eqref{Nu}. However, when the oscillation frequencies coincide, the interference term is given by \eqref{Nint} and is found to contribute constructive and destructive effects to different regions of the spectrum. We characterize the different ranges of the interference effect in \eqref{int w} and identify the necessary conditions on the input variables, the phase difference $\phi$ and values of $\lz$, $\uz$, and $\w_0$, that modify the constructive and destructive regions in the interference term. 

The total particle production is found for the interference term \eqref{creation rate int}, where we note an interesting feature of null particle production when $(1+\lz^2)\w_0/\uz\approx2.23$. This feature is present in other Casimir interference systems \cite{silva2015interference}, as this value corresponds to decoupling of the mirror from the field and can be related to the Robin boundary condition (with $\lz=1$ and $\gamma_0=2/\uz$ being the Robin parameter), which is associated with a strong inhibition of the particle production for $\gamma_0\w_0\approx2.23$ \cite{silva2016dynamical,mintz2006particle,rego2013inhibition}.

The difference between the spectrum produced on the left and right sides of the mirror, which encodes the asymmetry of the system, is quantified for the contributions from the motion of the mirror \eqref{DN moving dd} and the time-varying properties \eqref{DN properties}, along with the asymmetry from the interference \eqref{DN int}. Resonance behavior between the two different fluctuation sources is identified \eqref{DN resonence} and the dominant spectral contribution to the low and high frequency regimes are discussed; for low frequencies, the dominant contribution to the spectral difference comes from the fluctuations on the mirror's properties \eqref{low freq} and, for high frequencies, the spectrum from the motion of the mirror \eqref{high freq} is dominant. We find that the spectral difference can be written in a formula analogous to the wave intensity of the double-slit experiment \eqref{double slit form}, similar to the relationships found for the interference pattern of a cavity with moving oscillating walls \cite{ji1998interference} and for a moving mirror with a time-dependent Robin parameter \cite{silva2015interference}.

A model describing a stationary $\ddp$ mirror with two distinct, unspecified fluctuation sources, described by \eqref{mu 2 source}, is explored for the ADCE system. Here, enhancement of the asymmetry by means of increasing the spectral difference is achieved when both fluctuation sources are resonating at the same frequency. The enhanced spectral difference is presented in \eqref{enhance eps}. It is possible to tune this enhancement with the phase difference, the magnitude of the spectrum doubles when the system is fully in-phase and completely vanishes when the two frequencies are anti-aligned. We expand upon this model to include an arbitrary number of fluctuation sources \eqref{full enhance}. When all the fluctuation sources are in-phase, the spectrum is enhanced quadratically in the number of independent sources.

\begin{acknowledgments}
We wish to acknowledge Eric Davis, Patrick Brown, Jacob Matulevich, and Christian Brown for their helpful discussions and reviews.
\end{acknowledgments}


\bibliography{ref}

\end{document}